\documentclass{appolb}
\usepackage{graphicx}

\begin{document}
\def\Pom{\mathrm{I\!P}}
\def\Reg{\mathrm{I\!R}}

\title{
Light-by-light scattering in ultraperipheral heavy ion collisions
- new possibilities%
\thanks{Presented by A.S. at the Zakopane 2024 Conference
on Nuclear Physics,
Zakopane, Poland}%
}

\author{Antoni Szczurek
\address{Institute of Nuclear Physics PAN, PL-34-342 Krak\'ow, Poland 
and Rzesz\'ow University, PL-35-310 Rzesz\'ow, Poland}
\\[3mm]
Pawel Jucha
\address{Institute of Nuclear Physics PAN, PL-34-342 Krak\'ow, Poland}
}
\maketitle
\begin{abstract}
Light-by-light scattering is a relatively new area of experimental 
physics. Our recent, theoretical research shows that studying 
two photon measurements in regions with lower transverse momentum 
($p_{t,\gamma}$) and invariant mass ($M_{\gamma\gamma}$) 
allows us to observe not only the main contribution of photon 
scattering, known as fermionic loops but also mechanisms like 
the VDM-Regge (double-photon hadronic fluctuation). 
In addition, diphoton measurements 
at low diphoton masses are crucial for studies of light meson 
resonance contributions in $\gamma\gamma \to \gamma\gamma$ scattering. 
We also focus on the interference between different contributions. 
For future experiments with the ALICE FoCal and ALICE-3 detectors, 
we have calculated background contamination and have explored 
possibilities to minimize their impact.
\end{abstract}
 
\section{Introduction}

Light-by-light scattering, a quantum electrodynamic process where 
two photons interact and scatter off each other, is a phenomenon 
observed for the first time at the Large Hadron Collider (LHC) 
in 2017~\cite{Ref2}. Light-by-light scattering can be observed 
due to the intense electromagnetic fields generated by 
the ultrarelativistic nuclei. These fields can be considered 
as a flux of quasi-real photons that surround the ions. When 
two ions pass by each other, 
these electromagnetic fields can interact, enabling the 
light-by-light scattering. In ultraperipheral collisions
the process is essentially free of unwanted backgrounds.

Recent experiments at the LHC, conducted by the CMS~\cite{Ref3} 
and ATLAS~\cite{Ref4} collaborations, have successfully observed 
and measured light-by-light scattering in heavy ion collisions. 
However, both experiments use rather high thresholds for diphoton 
mass (5 GeV) and transverse momentum (2 GeV for CMS, and 2.5 GeV 
for ATLAS). The goal of our latest study \cite{Ref1} was to make 
predictions for future experiments with a lower threshold. 
Also, the study for different mechanisms, 
like VDM-Regge or light meson resonances, was conducted, 
searching for possibilities of their observation or even measurement.

\section{Sketch of the formalism}

\begin{figure}
    \centering
       a)\includegraphics[width=0.20\textwidth]{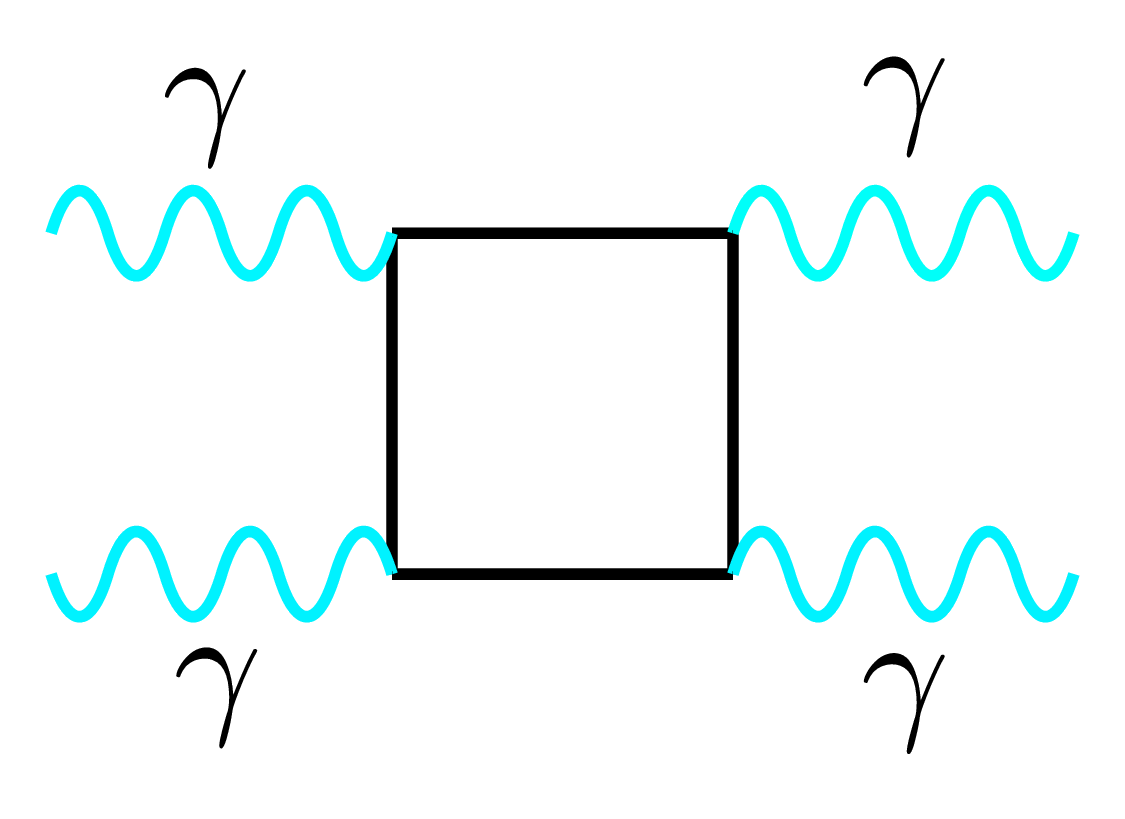}
       b)\includegraphics[width=0.20\textwidth]{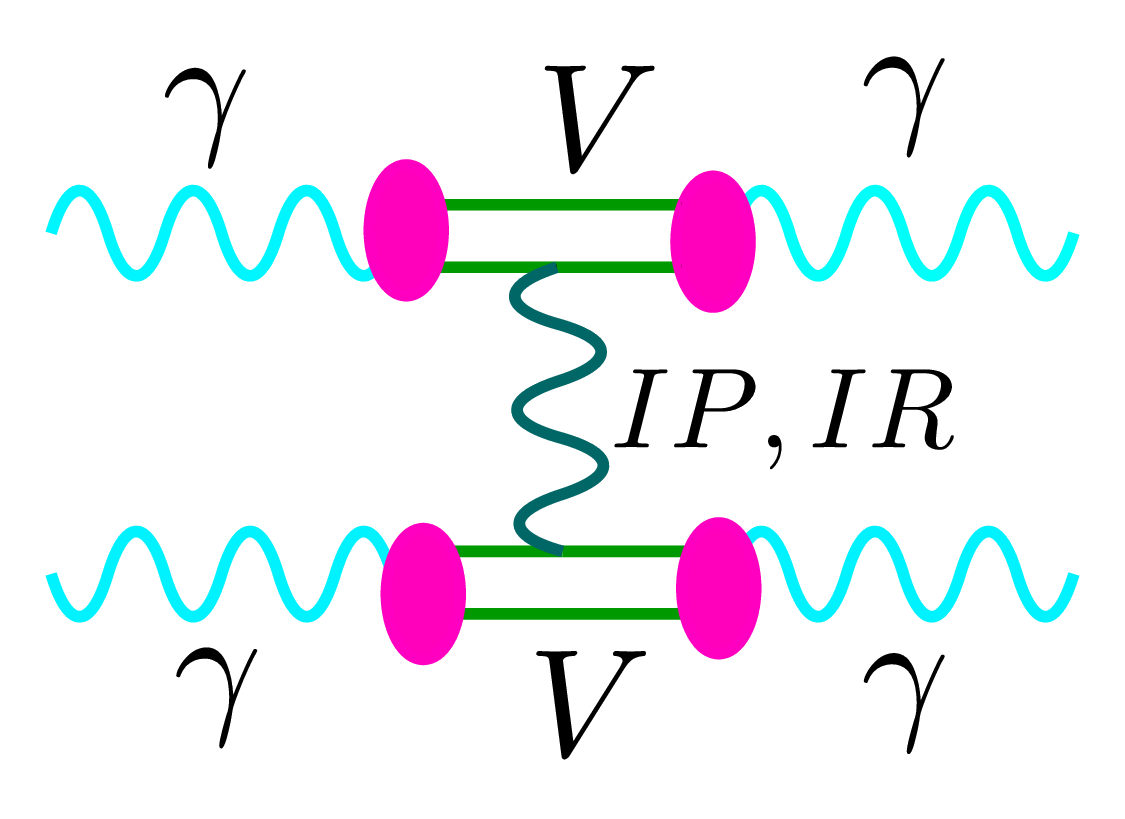}
       c)\includegraphics[width=0.20\textwidth]{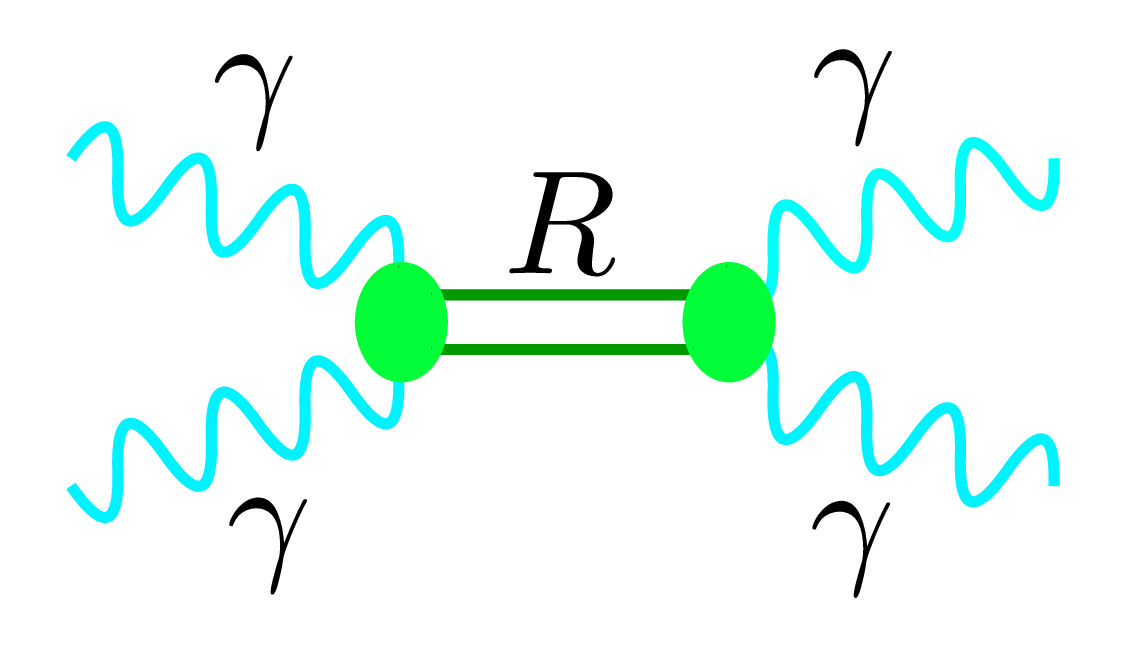}
       d)\includegraphics[width=0.20\textwidth]{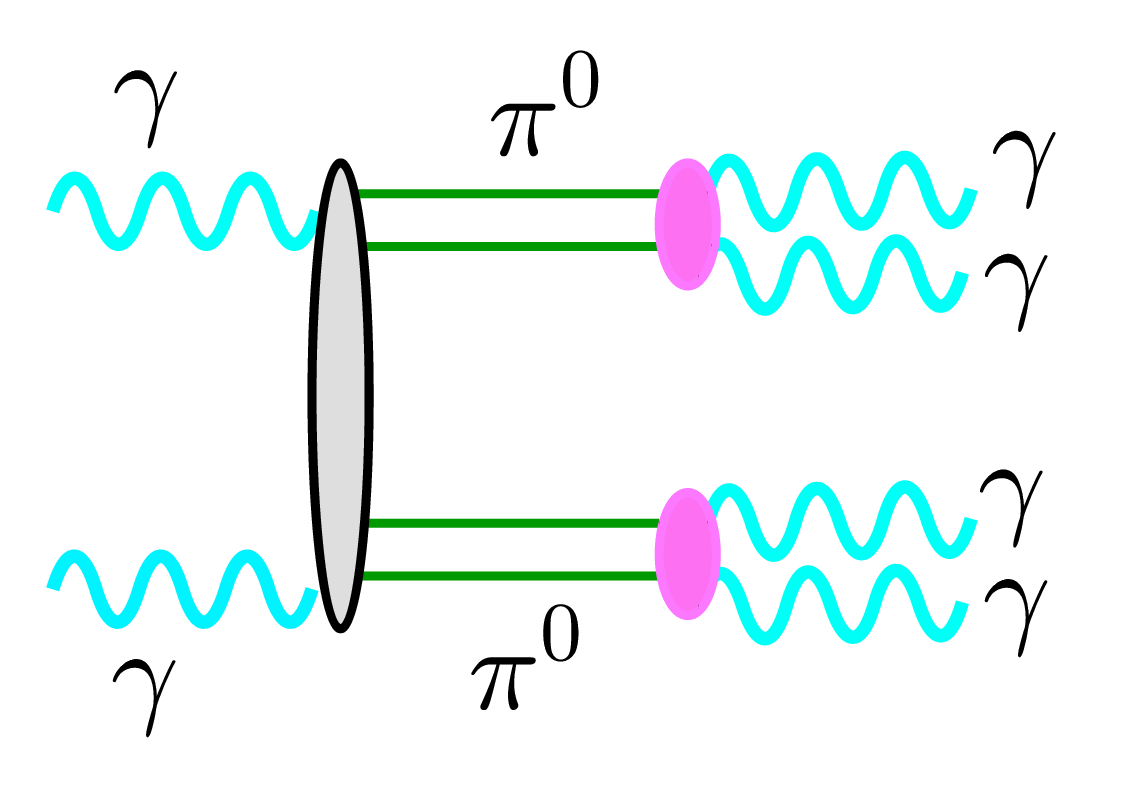}
    \caption{Feynman diagrams of the LbL scattering mechanisms: (a) fermionic loops, (b) VDM-Regge, (c) low-mass resonances in the s-channel, (d) two-$\pi^0$ background.}
    \label{fig:feynman_diag}
\end{figure}

\noindent The Equivalent Photon Approximation \cite{Ref11} relies 
on the knowledge about elementary cross section
as a function of diphoton mass $W_{\gamma\gamma}$ and distributions in 
the scattering angle, $z$~=~cos$\theta$.
In this paper various mechanisms of $\gamma\gamma \to \gamma\gamma$ 
process are taken into consideration.
The most common contribution is due to the four-vertex fermionic loops, 
so-called boxes, presented in Fig.~\ref{fig:feynman_diag}a. 
The calculation based on Feynman diagrams was carried out using 
FormCalc and LoopTools libraries
based on \textit{Mathematica} software \cite{Ref5}. 
To designate the cross sections for unpolarized photons 
16 photon helicity combinations of the amplitude must be added up 

 \begin{eqnarray}
     \sum_{{\lambda_1,\lambda_2,\lambda_3,\lambda_4}} \left|\mathcal{A}^{\gamma \gamma \rightarrow \gamma \gamma}_{\lambda_1\lambda_2 \to \lambda_3 \lambda_4}\right|^2 = 2\left|\mathcal{A}_{++++}\right|^2 + 2\left|\mathcal{A}_{+--+}\right|^2 \nonumber \\
     +2\left|\mathcal{A}_{+-+-}\right|^2 + 2\left|\mathcal{A}_{++--}\right|^2 
     +8\left|\mathcal{A}_{+-++}\right|^2 \;.
 \end{eqnarray}
Another  mechanism of light-by-light scattering is double-photon 
hadronic fluctuation (Fig.~\ref{fig:feynman_diag}b). The
mathematical description of the VDM-Regge mechanisms was presented in \cite{Ref6}:

\begin{eqnarray}
    {\cal M} &=& 
\Sigma_{i,j} C_i^2 C_j^2 
\left( 
  C_{\Pom} \left(\frac{s}{s_0} \right)^{\alpha_{\Pom}(t)-1} F(t)       
+ C_{\Reg} \left(\frac{s}{s_0} \right)^{\alpha_{\Reg}(t)-1} F(t) \right)
\; \nonumber \\
&+& 
\Sigma_{i,j} C_i^2 C_j^2 
\left( 
  C_{\Pom} \left(\frac{s}{s_0} \right)^{\alpha_{\Pom}(u)-1} F(u) 
+ C_{\Reg} \left(\frac{s}{s_0} \right)^{\alpha_{\Reg}(u)-1} F(u) \right) 
\; .
\end{eqnarray}

The light meson resonances (Fig~\ref{fig:feynman_diag}c) such as 
$\pi$, $\eta$ and $\eta'$ in $\gamma\gamma \to \gamma\gamma$ 
scattering are described using relativistic Breit-Wigner formula presented in \cite{Ref7}:

\begin{equation}
    \hspace{-0.5cm} \mathcal{M}_{\gamma\gamma\rightarrow R \rightarrow\gamma\gamma}(\lambda_1,\lambda_2) = 
    \frac{\sqrt{64\pi^2W^2_{\gamma\gamma}\Gamma_R^2Br^2(R\rightarrow\gamma\gamma)}}{\hat{s}-m_R^2-im_R\Gamma_R} \times \frac{1}{\sqrt{2\pi}}\delta_{\lambda_1-\lambda_2},
    \end{equation}

\noindent In the case of two photon measurement, the production of 
four photons from $\pi^0\pi^0$ decay 
(Fig.~\ref{fig:feynman_diag}d), where only two photons are observed 
in detectors constitutes the main background. Based on \cite{Ref8}, 
the idea how to reduce the background is discussed below.

\noindent The calculation of nuclear cross section for light-by-light
scattering is performed with the help of the equivalent photon
formula in the impact parameter space:

\begin{eqnarray}
    \frac{d\sigma(PbPb \to PbPb \gamma \gamma)}{dy_{\gamma_1}dy_{\gamma_2}dp_{t,\gamma}} = \nonumber \\
    \int \frac{d\sigma_{\gamma\gamma\to\gamma\gamma}(W_{\gamma\gamma})}{dz}
    N(\omega_1,b_1)N(\omega_2,b_2)S^2_{abs}(b) \noindent \\
    \times d^2b d\bar{b}_x d\bar{b}_y \frac{W_{\gamma\gamma}}{2} \frac{dW_{\gamma\gamma}dY_{\gamma\gamma}}{dy_{\gamma_1}dy_{\gamma_2}dp_{t,\gamma}} dz \;.
    \label{eq:tot_xsec}
\end{eqnarray}

\noindent Here the $\sigma_{\gamma \gamma \rightarrow \gamma \gamma}(W_{\gamma \gamma})$ is elementary cross section, $Y_{\gamma \gamma}$ is rapidity of outgoing two photons and $\overline{b}_x$, $\overline{b}_y$ are the components of the vector $(\vec{b}_1+\vec{b}_2)/2$, where $\vec{b} = \vec{b}_1 - \vec{b}_2.$ The $N(\omega_i,b_i)$ is photon flux, which is obtained from charge distribution in the nucleus. The $S^2(b)$ denotes the survival factor. In recent studies, the sharp edge formula ($S^2_{abs}(b) = \Theta(b-b_{max})$) was
replaced by the formula:

\begin{eqnarray}
    S^2_{abs}(b) = exp\left( -\sigma_{NN} T_{AA}(b) \right) \;,
    \label{eq:s2b}
\end{eqnarray}
%
where $\sigma_{NN}$ is the nucleon-nucleon interaction cross section, and $T_{AA}(b)$ is related to the so-called nuclear thickness function \cite{Ref1}.

\section{Results for $\gamma \gamma \to \gamma \gamma$ collisions}
\label{Sect.III}

\begin{figure}
    \centering
       a)\includegraphics[width=0.37\textwidth]{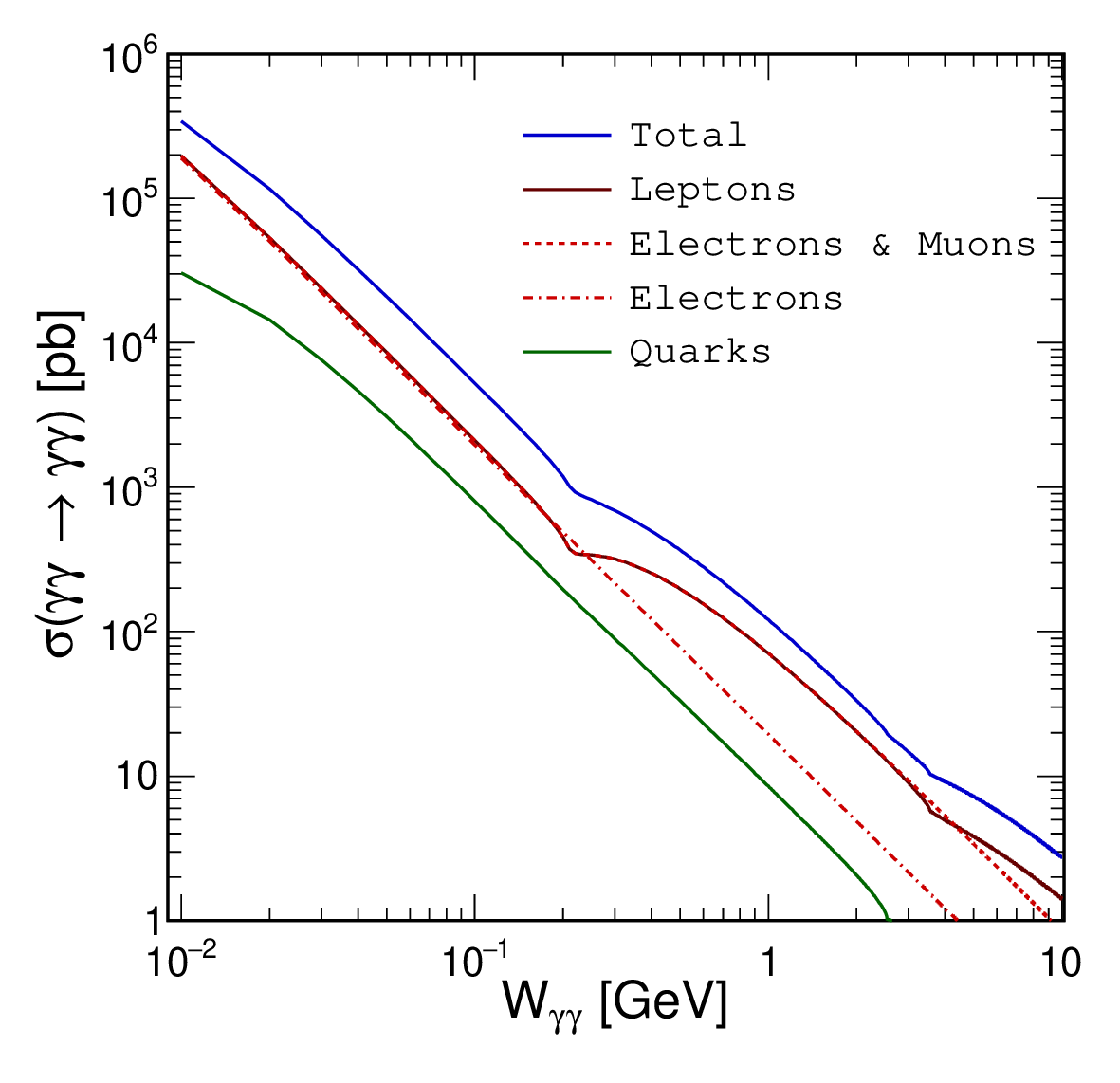}
       b)\includegraphics[width=0.37\textwidth]{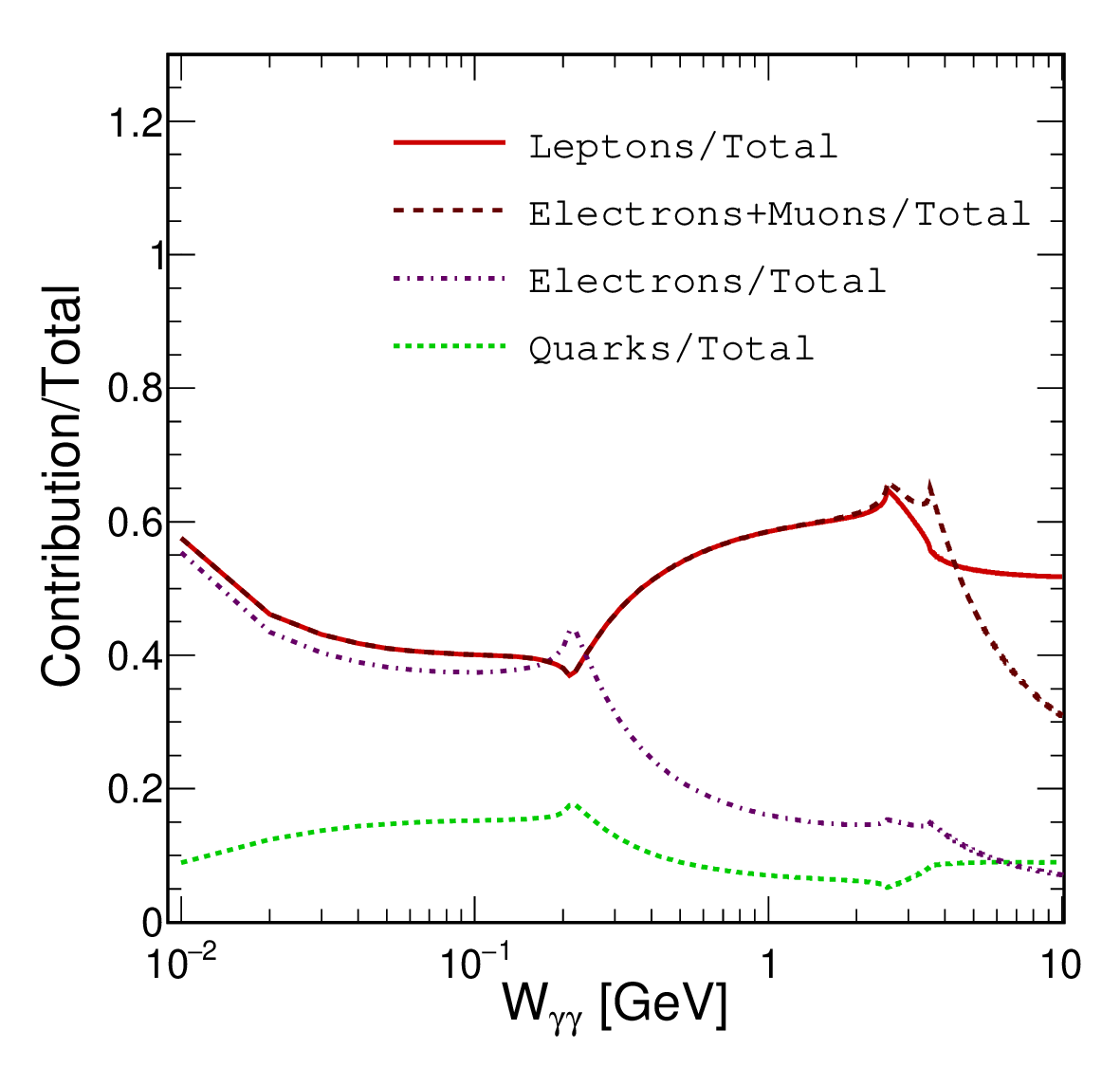}
    \caption{a) Elementary cross section (in pb) as a function of
                energy. The total cross section (blue solid line) is split
                somewhat artificially into quarks (green solid line),
                electrons (red dashed line), electrons and muons (red
                dotted line), and leptons (red solid line)
                contributions; (b) Ratio of each
                  contribution to a coherent sum of them: leptonic cross section divided 
by total cross section (red solid line), quarks (green dotted line), electrons (magenta dash-dotted line) and sum of electron and muon contributions (dashed dark-red line).}
    \label{fig:Elementary}
\end{figure}

\noindent Calculations of elementary cross section for light-by-light
scattering show that in diphoton invariant mass region below 1 GeV 
one can expect significant signal
from the loop contribution. Fig. \ref{fig:Elementary}a) 
displays separate contributions for different fermions in the loop. 
In Fig. \ref{fig:Elementary}b) the ratio of different contribution to 
the total cross section is presented. 
Here, the importance of interference is illustrated. 
The incoherent sum between quarkish and leptonic loops cannot 
sum up to 1. The impact of interference is estimated to be 
around 20\%.

\begin{figure}
    \centering
       \includegraphics[width=0.37\textwidth]{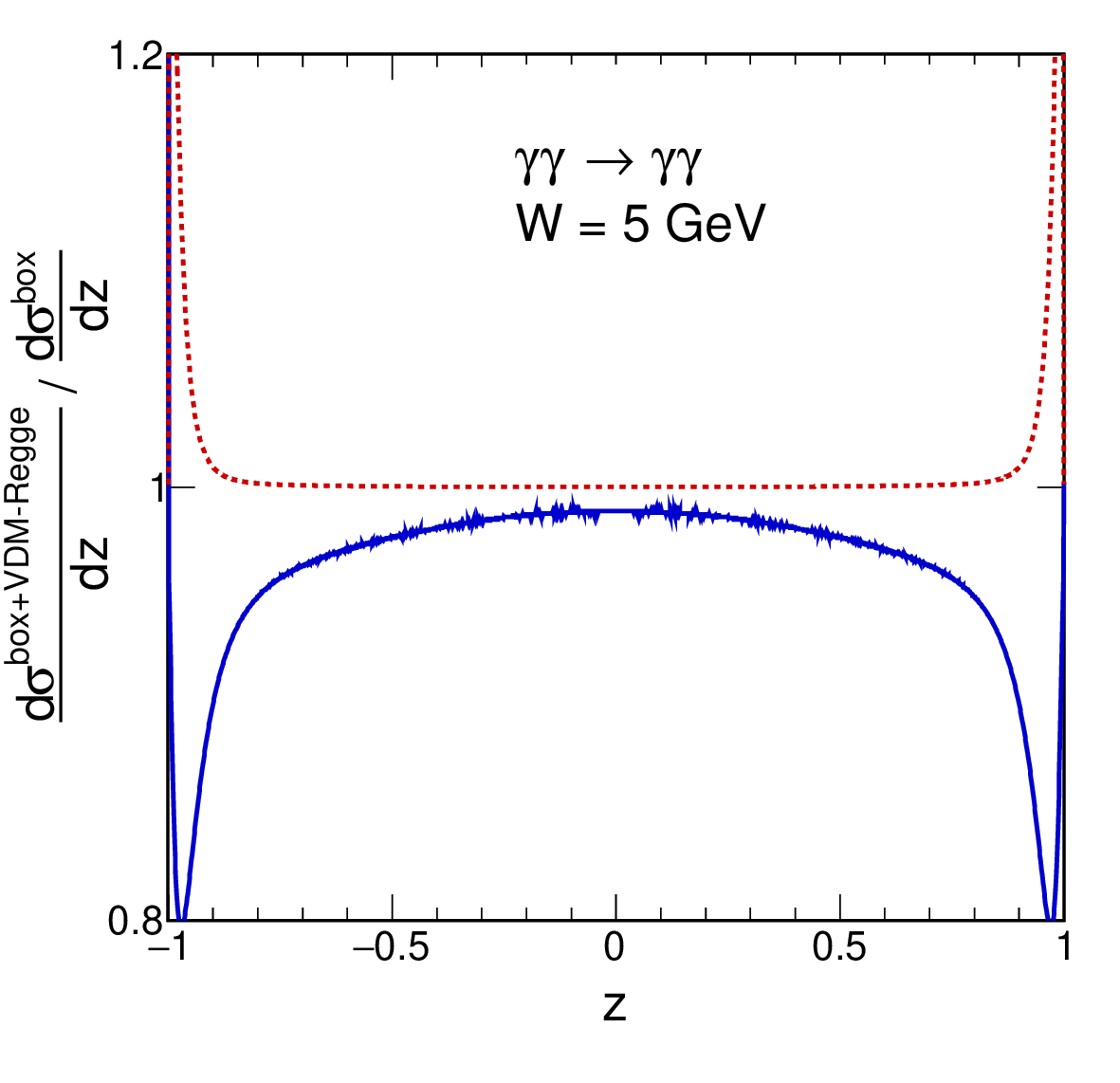}
    \caption{The ratio of the coherent (blue) and incoherent (red) sum 
		of the box and VDM-Regge contributions divided by the cross section for the box contribution alone for $W = 5$~GeV.}
    \label{fig:VDM_ratio}
\end{figure}

\noindent The analysis of interference of box and VDM-Regge 
mechanisms was also conducted.
In Fig. \ref{fig:VDM_ratio} we show the plots of ratios of the sum 
of two contributions (VDM-Regge, boxes) to the box contributions
as a function of $z$~=~cos$\theta$ variable.
The role of interference is clearly visible for $|z|$~$\to$~1. The
smaller the scattering angle, the greater the impact of VDM-Regge
contribution on the cross section. We found that the effect 
of interference is destructive.

\section{Results for nuclear collisions}
\label{Sect.IV}

\begin{figure}
    \centering
       a)\includegraphics[width=0.37\textwidth]{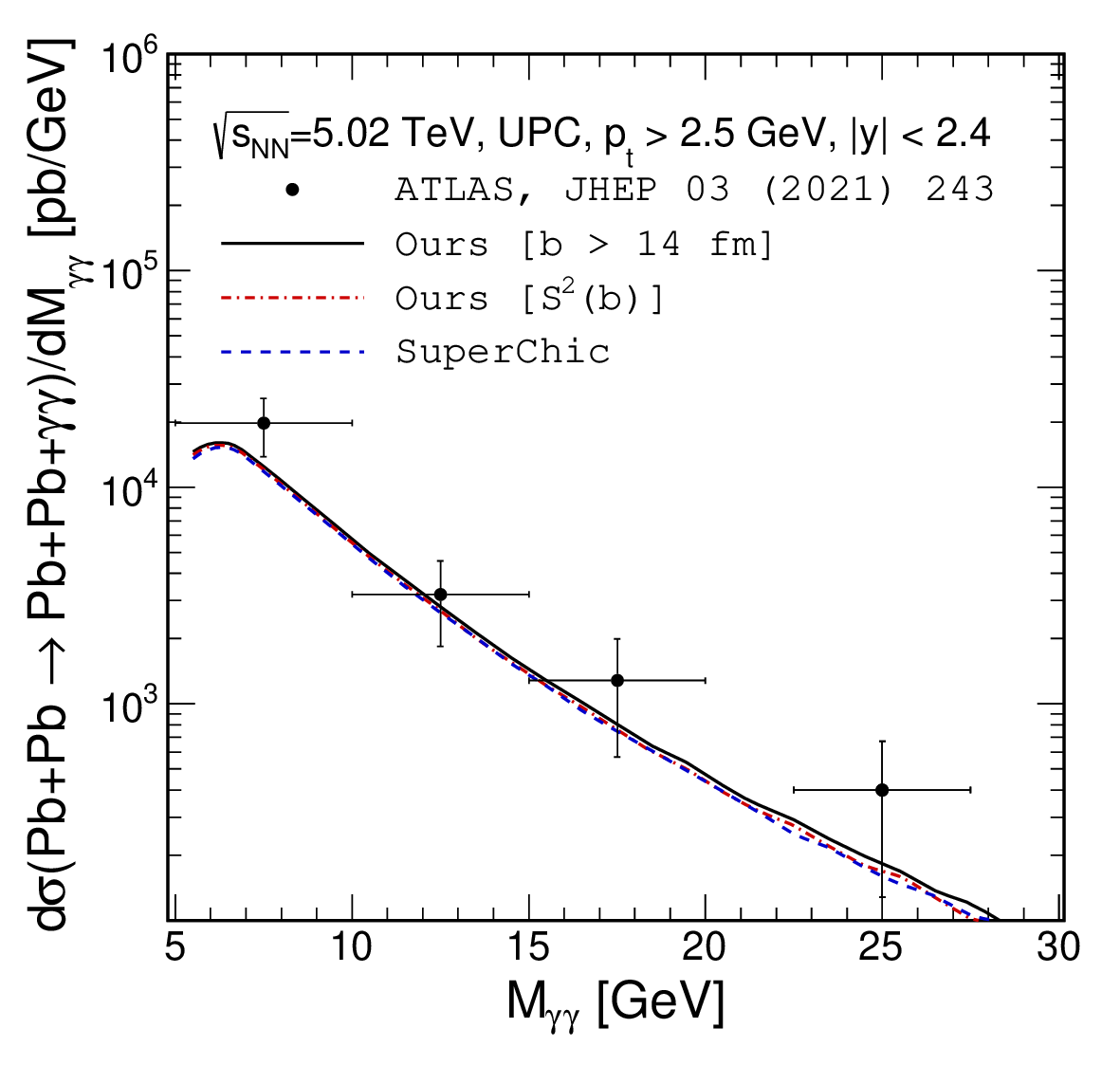}
       b)\includegraphics[width=0.37\textwidth]{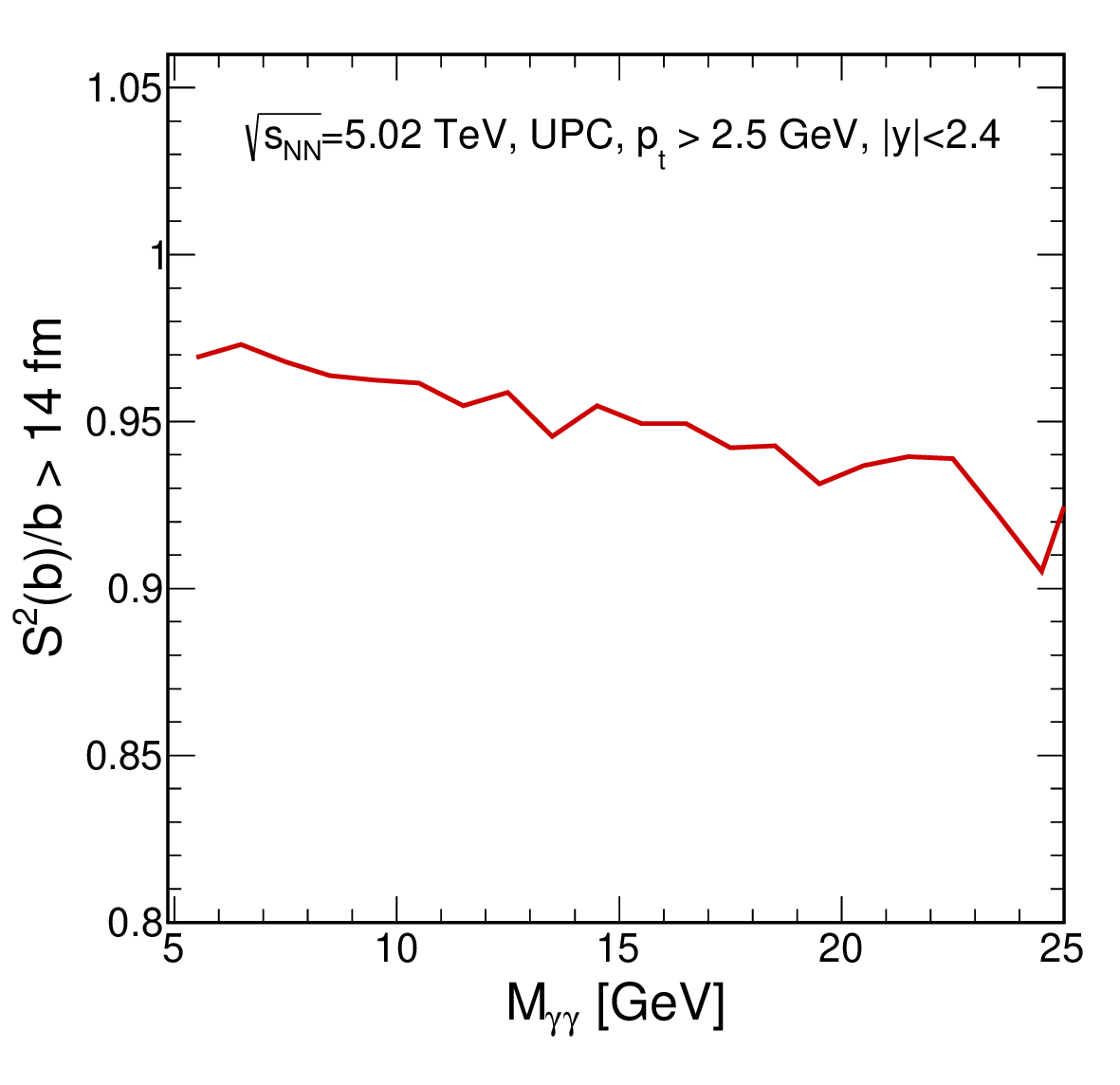}
    \caption{Differential cross section as a function
          of diphoton invariant mass at $\sqrt{s_{NN}}=5.02$~TeV. (a)
          The ATLAS experimental data are shown together with 
          theoretical results including a sharp cut on impact 
          parameter ($b>14$~fm
          (solid black line) and smooth nuclear absorption factor
          $S^2(b)$ (dash-dotted red line). Results obtained from 
          SuperChic \cite{HarlandLang2020veo} is also shown for comparison.
          (b) distributions calculated within our approach with 
          sharp and smooth cut-off on impact parameter.
          The ratio of cross sections for smooth to sharp cut-off
          is shown in panel b).}
    \label{fig:s2b}
\end{figure}

\noindent The approach with smooth survival factor (Eq. (\ref{eq:s2b}))
was implemented to calculation for the ATLAS data. 
In Fig. \ref{fig:s2b}a) the distribution of diphoton invariant
mass for ATLAS kinematical cuts with sharp edge of nucleus and the survival factor based on Eq.~(\ref{eq:s2b}) is presented. All theoretical approaches, including SuperChic, underestimate the cross section in low-mass region.
In Fig. \ref{fig:s2b}b) ratio of results for both our approaches 
for impact parameter cutoff is shown. The difference is changing between 4\% for smaller values of diphoton mass up to 10\% for higher masses.

\begin{figure}
    \centering
       \includegraphics[width=0.37\textwidth]{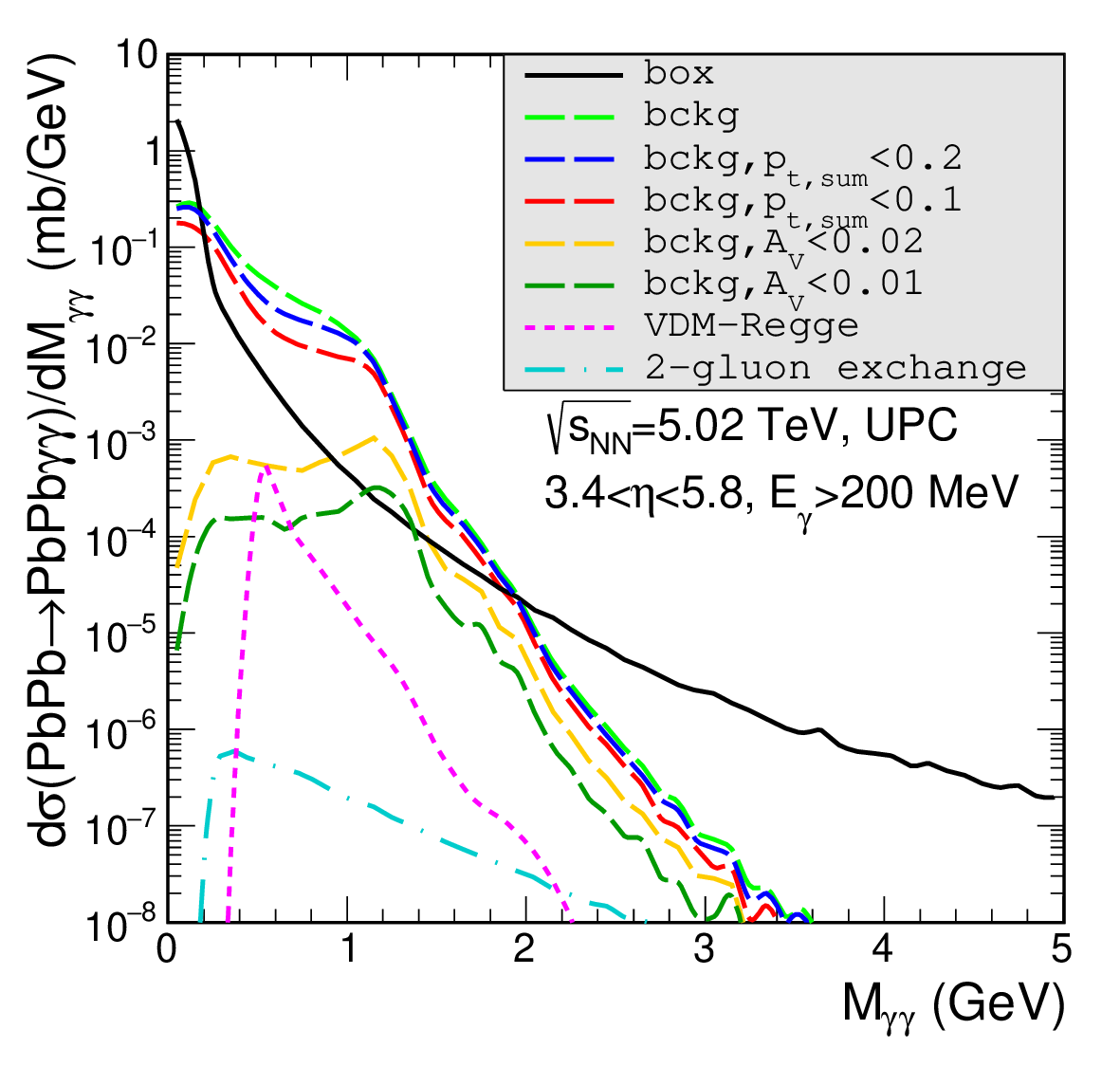}
    \caption{Invariant mass distribution for the nuclear process.
Shown are predictions for the future FoCal acceptance 
$E_{t,\gamma}>200$~MeV and $3.4<y_{\gamma_{1/2}}<5.8$.}
    \label{fig:focal}
\end{figure}

\noindent A planned forward detector for ALICE experiment - FoCal, 
will start collecting data since the Run 4 \cite{Ref9}. 
Kinematical cuts for photons were applied, to take advantage of the
lower threshold for diphoton invariant mass ($M_{\gamma\gamma} > 500$
MeV), and the accessibility of larger photon rapidity interval 
($3.4<y_{\gamma_{1/2}}<5.8 $). The distribution of diphoton mass is
presented in Fig. \ref{fig:focal}. In this low-mass region, the
background from the $\pi^0\pi^0$ production become significant. However,
using isotropic 
distribution of photons in each $\pi^0$ decay
and imposing condition of two photons in the experimental acceptance
one can suppress the background. Hence, a cut for vector asymmetry 
$A_V = |\vec{p}_{t,1} - \vec{p}_{t,2}| / |\vec{p}_{t,1} +
\vec{p}_{t,2}|$ was proposed. The implied cut notably reduces 
the measured cross section for the $\pi^0\pi^0$ background.

\begin{figure}[h]
    \centering
       \includegraphics[width=0.37\textwidth]{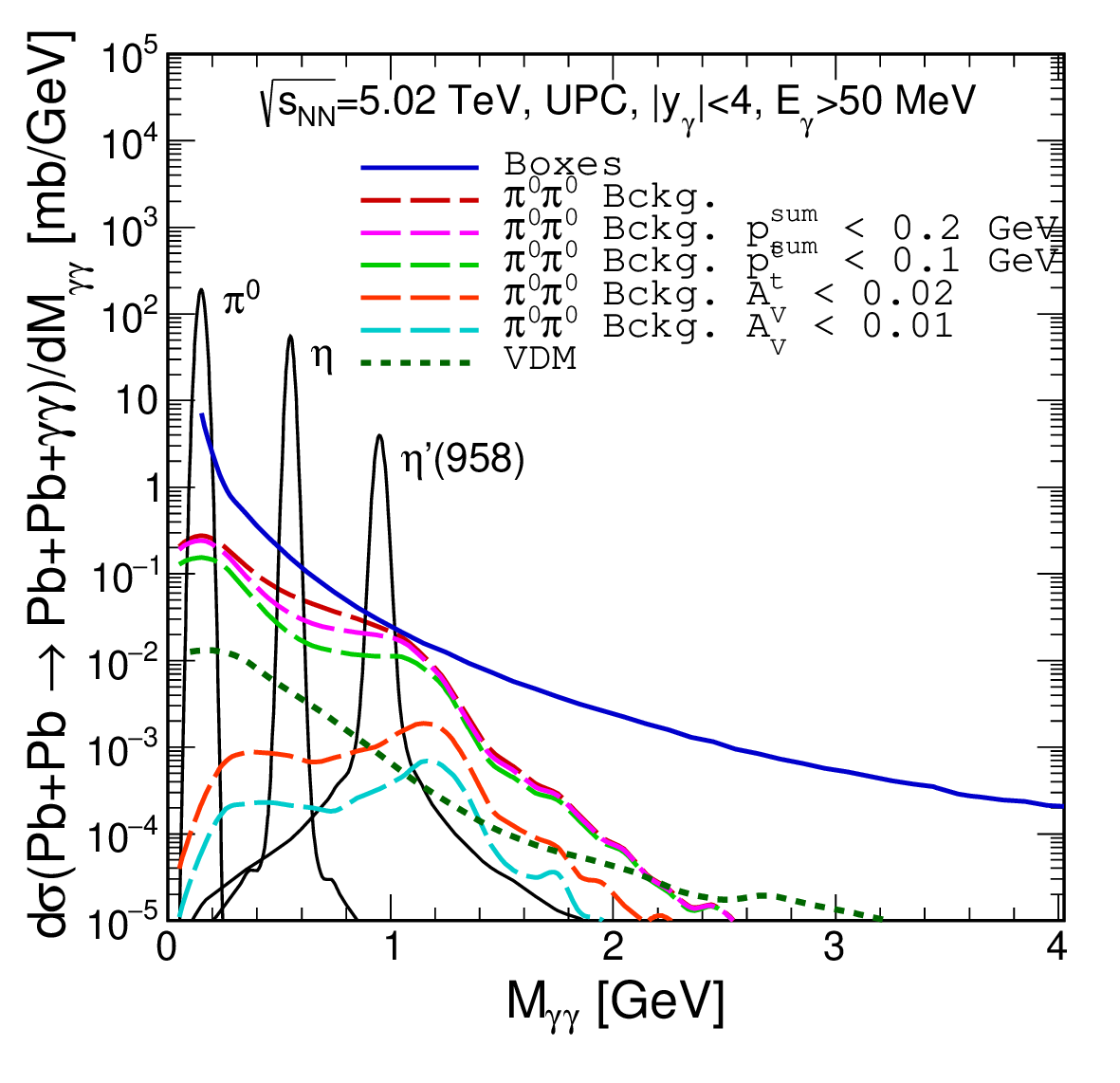}
    \caption{ Diphoton invariant mass distribution for ALICE-3, i.e. 
 $y_{\gamma_{1/2}} \in (-4,4)$ and photon energy E$_{\gamma}>50$~MeV. Here the black solid line represents the light meson resonances, the blue solid line relates to the box contribution, the dotted line to the VDM-Regge component and the dashed lines are for double-$\pi^0$ background contribution.}
    \label{fig:ALICE3}
\end{figure}

\noindent From theoretical perspective, ALICE-3 experiment 
presents further opportunities for observation of light-by-light 
scattering in a new kinematical region.
The calculation of cross
sections were prepared based on the characteristics of the detector 
including kinematical limitations \cite{Ref10}. 
Fig. \ref{fig:ALICE3} shows the diphoton mass distribution in
the assumed kinematical range. In the diphoton mass region below 1 GeV
the most significant signal comes from the mesonic resonances. 
The peaks from $\pi^0$, $\eta$ and $\eta'$
overachieve even the box contribution, which will allow to observe 
this contributions for the first time. 
Here, the impact of the background can also be reduced using cuts 
on so-called vector asymmetry.

\begin{figure}
    \centering
       a)\includegraphics[width=0.37\textwidth]{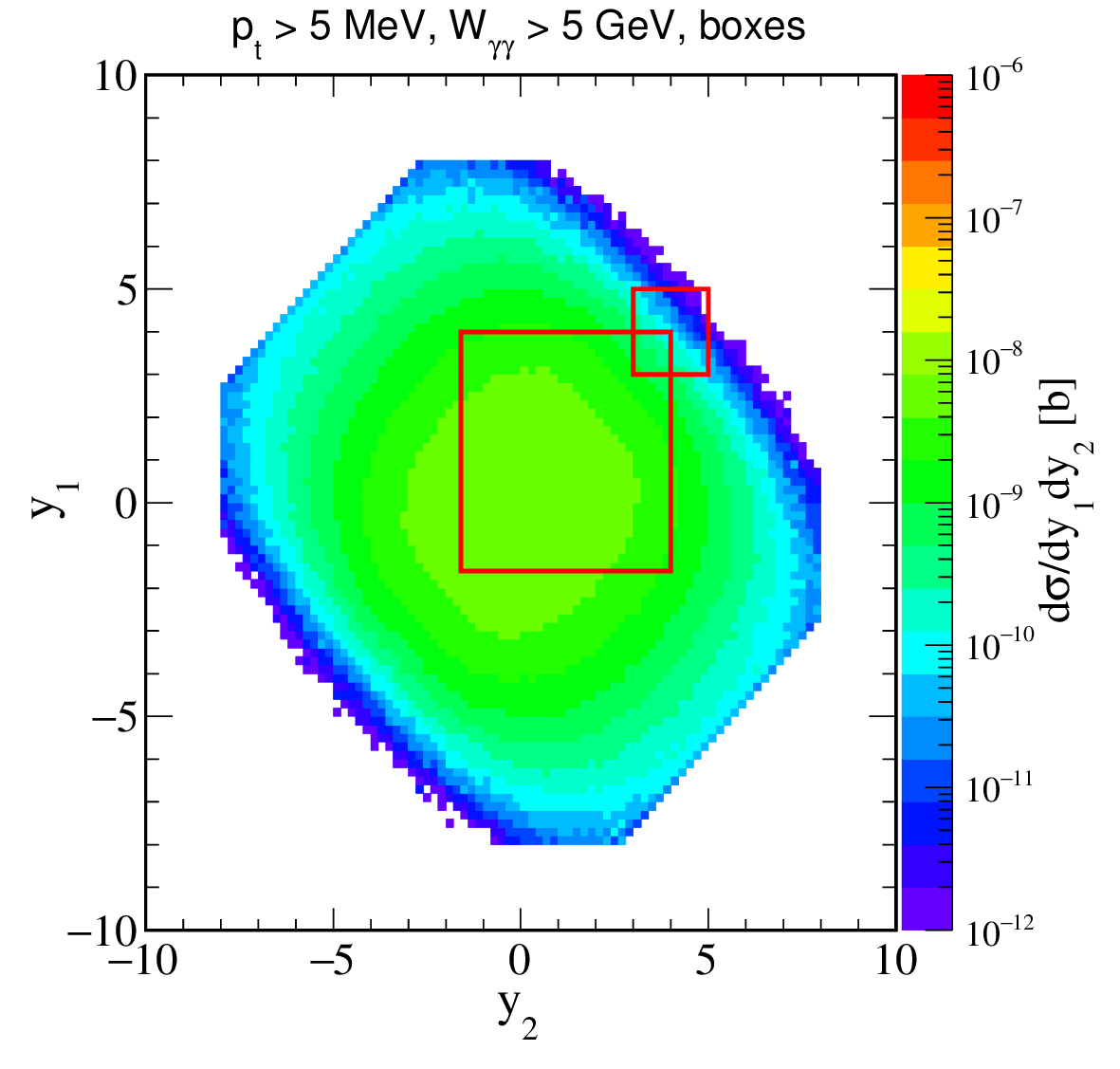}
       b)\includegraphics[width=0.37\textwidth]{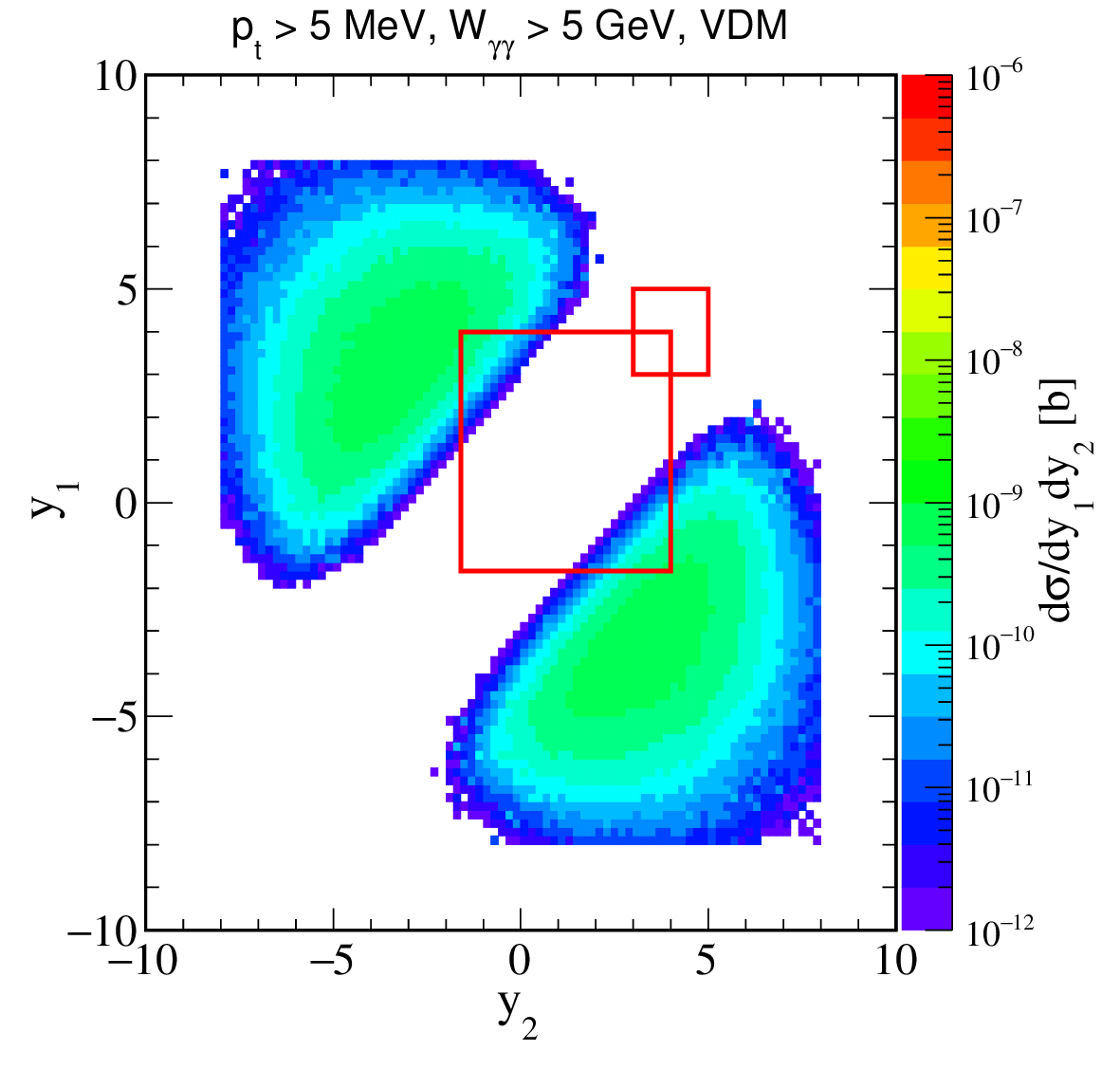}
    \caption{Distribution in ($y_{\gamma_1}$, $y_{\gamma_2}$) in b 
for transverse momenta $p_{t, \gamma}>5$~MeV, di-photon invariant 
mass $M_{\gamma\gamma}>5$~GeV. (a) Boxes, (b) VDM-Regge mechanism.}
    \label{fig:rapidity}
\end{figure}

\noindent The most challenging task would be a measurement of the
VDM-Regge mechanism. Presented in Fig. \ref{fig:VDM_ratio} 
cross section distribution shows that the scattered photons in 
the VDM-Regge process
 are mainly forward/backward. This fact implies that the dedicated 
experiment has to have a broad range of photon rapidities. 
Fig.~\ref{fig:rapidity} presents a two-dimensional rapidity 
distributions of a) box continuum, b) VDM-Regge mechanism. 
The red squares represents the cover of rapidity
ranges of the ALICE-3 barrel ($-1.6< y_{\gamma_{1/2}} < 4$) and forward
($3 < y_{\gamma_{1/2}} < 5$) detectors. This plots reveals that current
and planned experiments avoid the region of interest.

\begin{figure}
    \centering
       \includegraphics[width=0.37\textwidth]{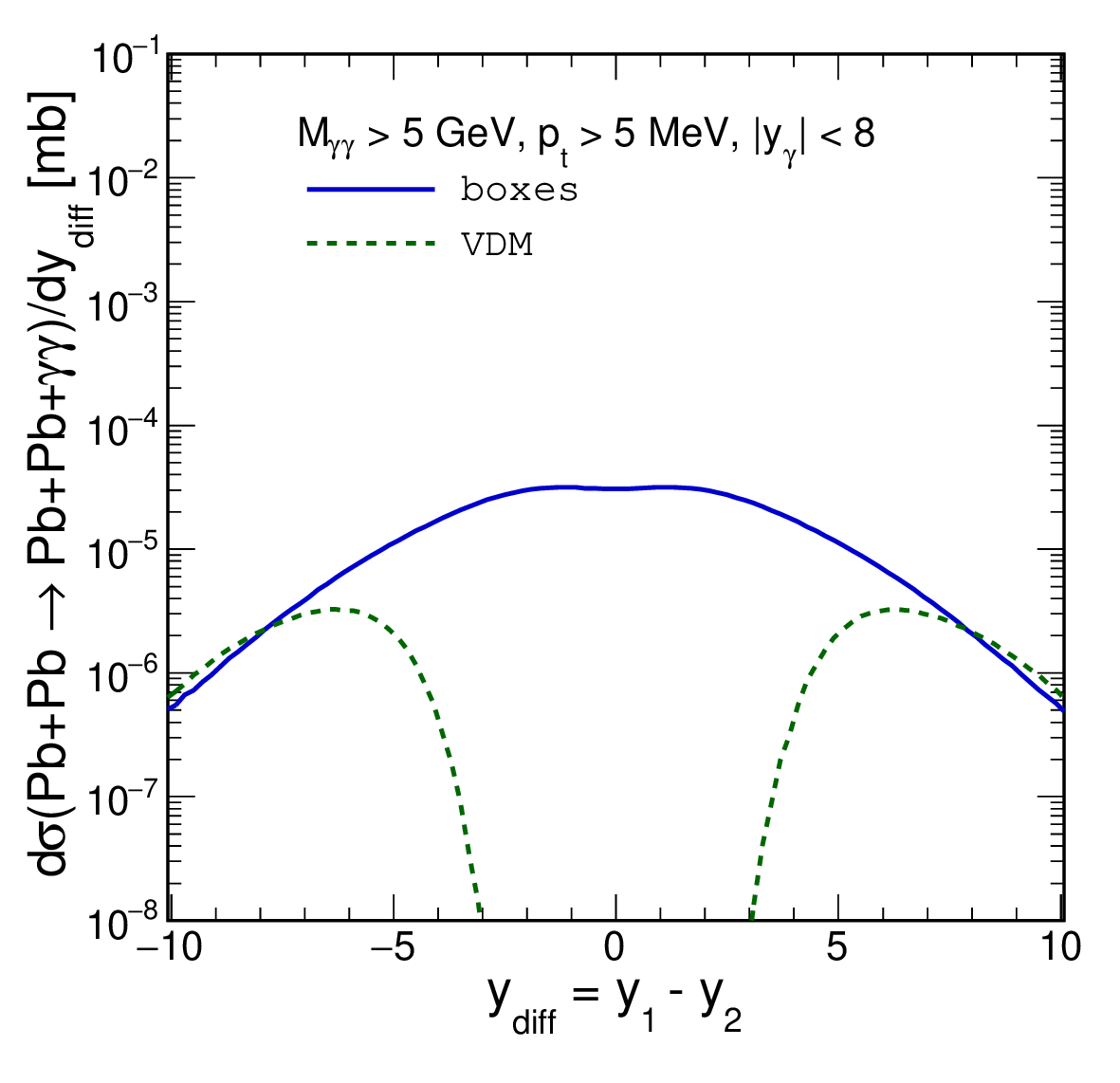}
    \caption{Distribution in $y_{diff}$ for light-by-light scattering 
processes in PbPb$\to$PbPb$\gamma\gamma$. Here the transverse 
momentum cut is equal to 5~MeV only. The blue solid line relates to the
boxes, and the green dotted line to the VDM-Regge contribution. Here the
range of measured diphoton invariant mass is 
$M_{\gamma \gamma}> 5$~GeV.}
    \label{fig:y-diff}
\end{figure}

\noindent However, the introduction of a new variable - the rapidity difference, may facilitates this observation.
Fig.~\ref{fig:y-diff} displays the distribution of the fermionic loops and the VDM-Regge mechanism in
the mentioned variable. For the higher values of $|y_{diff}|$ the chances for observing the VDM-Regge mechanisms
increase.

\section{Conclusion}

\noindent The future development of new photons detectors opens up 
new opportunities for 
light-by-light scattering measurements. The lower threshold of diphoton mass
should enable observation of light mesonic resonances. In recent studies, also the
method of reduction of the background from the $\pi^0\pi^0$ decays 
was discussed.
The goal of future experimental measurements should be the observation of the
VDM-Regge mechanism, not observed so far. However, required range of measured photon rapidity is difficult
to achieved. Also, the extinguishing character of the interference 
of the fermionic loops and the VDM-Regge mechanism generates 
additional difficulties.\\

\vspace{0.5cm}

{\bf Acknowledgements}

A.S. is indebted to Mariola K{\l}usek-Gawenda
for collaboration on the issues presented here.

%


\begin{thebibliography}{}


\bibitem{Ref2}
M. Aaboud and others (ATLAS Collaboration), Nature Phys. \textbf{13}, 9, 852-858 (2017).

\bibitem{Ref3}
A. M. Sirunyan and others (CMS Collaboration), Phys. Lett. B \textbf{797}, 134826 (2019).

\bibitem{Ref4}
G. Aad and others (ATLAS Collaboration), Phys. Rev. Lett. \textbf{123}, 052001 (2019).

\bibitem{Ref1}
P. Jucha, M. Kłusek-Gawenda, A. Szczurek, Phys. Rev. D \textbf{109}, 014004 (2024).

\bibitem{Ref11}
M. Kłusek-Gawenda, A. Szczurek, Phys. Rev. C \textbf{82}, 014904 (2010).

\bibitem{Ref5}
T. Hahn and M. P{\'{e}}rez-Victoria, Computer Physics Communications \textbf{118}, 2-3, 153-165 (1999).

\bibitem{Ref6}
M. Kłusek-Gawenda, P. Lebiedowicz, and A. Szczurek, Phys. Rev. C \textbf{93}, 044907 (2016).


\bibitem{Ref7}
M. Kłusek-Gawenda, R. McNulty, R. Schicker, and A. Szczurek, Phys. Rev. D \textbf{99}, 093013 (2019).

\bibitem{Ref8}
M. Kłusek-Gawenda, and A. Szczurek, Phys. Rev. C \textbf{87}, 054908 (2013)

\bibitem{HarlandLang2020veo}
L. A. Harland-Lang, M. Tasevsky,  V. A. Khoze, M. G. Ryskin, Eur. Phys. J. C \textbf{80}, 10, 925 (2020).

\bibitem{Ref9}
C. Loizides, W. Riegler et al. (ALICE Collaboration), CERN Document Server (2020), arXiv:2211.0249

\bibitem{Ref10}
L. Musa, W. Riegler (ALICE collaboration), , CERN Document Server (2022), arXiv:2211.0249



\end{thebibliography}

\end{document}